\documentclass[A4paper]{IEEEtran}
\usepackage[left=0.52in, right=0.52in, bottom=0.9in, top=19mm]{geometry}
\pdfoutput=1
\IEEEoverridecommandlockouts
\usepackage{cite}
\usepackage{amsmath,amssymb,amsfonts}
\usepackage{amscd}
\usepackage{epsfig}
\usepackage{psfrag}
\usepackage{rotating}
\usepackage{float}
\usepackage{url}
\usepackage{tikz}
\usepackage{pgfplots}
\usepackage{mathtools}
\usepackage{algorithmic}
\usepackage{graphicx}
\usepackage{array}
\usepackage{siunitx}

\DeclareSIUnit{\belmilliwatt}{Bm}
\DeclareSIUnit{\belsquaremeter}{Bsm}
\DeclareSIUnit{\bit}{bits}
\DeclareMathOperator*{\argmax}{arg\,max}
\DeclareMathOperator*{\argmin}{arg\,min}
\graphicspath{ {./figures/} }
\usepackage{textcomp}
\usepackage{xcolor}
\usepackage[nolist,nohyperlinks]{acronym}
\long\def\comment#1{}

\newfont{\bbb}{msbm10 scaled 700}

\newfont{\bb}{msbm10 scaled 1100}

\newcommand{\CC}{\mbox{\bb C}}

\newcommand{\RR}{\mbox{\bb R}}

\newcommand{\EE}{\mbox{\bb E}}

\newcommand{\av}{{\bf a}}
\newcommand{\bv}{{\bf b}}
\newcommand{\cv}{{\bf c}}

\newcommand{\fv}{{\bf f}}

\newcommand{\nv}{{\bf n}}

\newcommand{\sv}{{\bf s}}

\newcommand{\uv}{{\bf u}}
\newcommand{\wv}{{\bf w}}

\newcommand{\xv}{{\bf x}}
\newcommand{\yv}{{\bf y}}

\newcommand{\zerov}{{\bf 0}}

\newcommand{\Dm}{{\bf D}}

\newcommand{\Gm}{{\bf G}}
\newcommand{\Hm}{{\bf H}}
\newcommand{\Id}{{\bf I}}

\newcommand{\Rm}{{\bf R}}

\newcommand{\Tm}{{\bf T}}
\newcommand{\Um}{{\bf U}}

\newcommand{\Vm}{{\bf V}}
\newcommand{\Xm}{{\bf X}}
\newcommand{\Ym}{{\bf Y}}

\newcommand{\Cc}{{\cal C}}

\newcommand{\Nc}{{\cal N}}

\newcommand{\Xc}{{\cal X}}

\newcommand{\xiv}{\hbox{\boldmath$\xi$}}

\newcommand{\Phim}{\hbox{\boldmath$\Phi$}}

\newcommand{\diag}{{\hbox{diag}}}
\renewcommand{\det}{{\hbox{det}}}

\renewcommand{\Re}{{\rm Re}}
\renewcommand{\Im}{{\rm Im}}

\newcommand{\herm}{{\sf H}}
\newcommand{\transp}{{\sf T}}

\newcommand{\T}{{\scriptscriptstyle\mathsf{T}}}

\newcommand{\rsf}{\boldsymbol{r}}
\newcommand{\wsf}{\boldsymbol{w}}
\newcommand{\xsf}{\boldsymbol{x}}
\newcommand{\ysf}{\boldsymbol{y}}
\newcommand{\rect}{{\sf rect}}

\newcommand{\Na}{N_{\rm a}}
\newcommand{\Nrf}{N_{\rm rf}}

\newcommand{\Lrf}{L_{\rm rf}}

\newcommand{\Pav}{P_{\rm av}}

\newcommand{\La}{L_{\rm a}}
\newcommand{\Tcp}{T_{\rm cp}}

\usepackage{hyperref}
\hypersetup{
    bookmarks=true,         
    unicode=false,          
    pdftoolbar=true,        
    pdfmenubar=true,        
    pdffitwindow=false,     
    pdfstartview={FitH},    
    pdfnewwindow=true,      
    colorlinks=true,       
    linkcolor=red,          
    citecolor=green,        
    filecolor=magenta,      
    urlcolor=cyan           
}

\def\BibTeX{{\rm B\kern-.05em{\sc i\kern-.025em b}\kern-.08em
    T\kern-.1667em\lower.7ex\hbox{E}\kern-.125emX}}
\begin{document}

\title{Beam Alignment with an Intelligent Reflecting Surface for Integrated Sensing and Communication\\
}

	\author{\IEEEauthorblockN{
	        Florian Muhr\IEEEauthorrefmark{1},
	        Lorenzo Zaniboni\IEEEauthorrefmark{1},
			Saeid K. Dehkordi\IEEEauthorrefmark{2},
		    Fernando Pedraza\IEEEauthorrefmark{2},
			Giuseppe Caire\IEEEauthorrefmark{2}}
		    \IEEEauthorblockA{\\
		    \IEEEauthorrefmark{1}Technical University of Munich, Germany\\
			\IEEEauthorrefmark{2}Technical University of Berlin, Germany\\\
			Emails: \{f.muhr, lorenzo.zaniboni\}@tum.de, \{s.khalilidehkordi, f.pedrazanieto, caire\}@tu-berlin.de}}

\maketitle
\begin{acronym}
    \acro{OFDM}{orthogonal frequency-division multiplexing }
    \acro{Tx}{transmitter }
    \acro{IRS}{intelligent reflecting surface}
	\acro{AWGN}{additive white Gaussian noise}
	\acro{MIMO}{multiple-input multiple-output}
	\acro{ULA}{uniform linear array}
	\acro{CRLB}{Cram\'er-Rao lower bound}
	\acro{SNR}{signal-to-noise ratio}
	\acro{mmWave}{millimeter wave}
	\acro{ML}{maximum likelihood}
	\acro{BS}{base station}
	\acro{UE}{user equipment}
	\acro{HDA}{hybrid digital-analog} 
	\acro{ISAC}{integrated sensing and communication}
	\acro{RF}{radio frequency}
	\acro{BA}{beam alignment}
	\acro{AoA}{angle of arrival}
	\acro{AoD}{angle of departure}
	\acro{LOS}{line-of-sight}
	\acro{CP}{cyclic prefix}
	\acro{ISI}{inter-symbol interference}
	\acro{BF}{beamforming}
	\acro{DFT}{discrete Fourier transform}
	\acro{RCS}{radar cross-section}
	\acro{MMLE}{multi-slot maximum likelihood estimation}
	\acro{HIRS}{hybrid-intelligent reflective surface}  
	\acro{DL}{downlink}
	\acro{UL}{uplink}
	\acro{RMSE}{root mean square error}
\end{acronym}

\begin{abstract}
In a typical communication system, in order to maintain a desired \ac{SNR} level, initial \ac{BA} must be established prior to data transmission. In a setup where a \ac{BS} \ac{Tx} sends data via a digitally modulated waveform, we propose an \ac{UE} enhanced with an \ac{HIRS} to aid beam alignment. A novel multi-slot estimation scheme is developed that alleviates the restrictions imposed by the \ac{HDA} architecture of the \ac{HIRS} and the \ac{BS}. To demonstrate the effectiveness of the proposed \ac{BA} scheme, we  derive the \ac{CRLB} of the parameter estimation scheme and provide numerical results.
\end{abstract}



\section{Introduction}
Integrated sensing and communication is emerging as a key component of  beyond-5G and 6G wireless systems\cite{9705498}. The increasing demand for higher data rates has led to considering \ac{mmWave} communications with its large frequency bandwidths. These frequencies exhibit high isotropic path loss so that a large beamforming gain is required, which can be achieved by using large antenna arrays and aligning the directional beams of the \ac{UE} and \ac{BS}. However, sampling broadband signals of many antennas is in general expensive, which motivates the use of \ac{HDA} architectures \cite{HDA_Sohrabi} at the \ac{BS} and \ac{UE} to reduce hardware cost. We propose to equip a \ac{UE} with a hybrid-intelligent reflective surface (\ac{HIRS}) to aid beam alignment (\ac{BA}). In such a setup, the \ac{IRS} array is physically mounted on the \ac{UE} and enables \ac{ISAC}.
The majority of recent studies have focused on positioning intelligent surfaces between the \ac{BS} and \ac{UE} in a fixed manner, where they serve as configurable reflectors to modify the propagation environment. The main objective is to have the \ac{IRS} either extend the range, increase the rank of the channel matrix \cite{ozdogan}, or enhance the (radar-) sensing capability \cite{Zahra}. \ac{UE}s equipped with \ac{IRS} have recently been studied in \cite{IRS_JSAC}, where the authors suggest to install large \ac{IRS} arrays on vehicles to improve the sensing of automotive users.
The work of \cite{Schober} has investigated the use of \textit{Simultaneously Transmitting and Reflecting} \ac{IRS}, where the incident wireless signal is divided into transmitted and reflected signals passing into both sides of the space surrounding the surface. The authors of \cite{alexandropoulos2021hybrid} introduce the concept of \ac{HIRS}, which enables metasurfaces to reflect the impinging signal in a controllable manner, while simultaneously sensing a portion of it. In this work we also adopt such an \ac{IRS} architecture. Note that this architecture differs from that in \cite{Schober} in that the former re-transmits a portion of the impinging wave via another set of antenna elements.

The main contribution of this work is to present a scheme where the \ac{BS} and a \textit{mobile} \ac{UE} that is equipped with such \ac{HIRS}, perform parameter estimation at each end. The \ac{HDA} architecture has a limited number of RF chains at both entities which prohibits conventional \ac{MIMO} processing and calls for design of RF domain reduction matrices. Taking this into account, we develop a multi-slot scheme where the reduction matrices achieve a trade-off between exploring the beam-space and high beamforming gain. The contributions of this work are summarized below:
\begin{itemize}
    \item We propose to use an \ac{HIRS} equipped \ac{UE} to assist the initial \ac{BA} procedure for highly directional beamforming applications.
    \item To meet the constraints of the \ac{HDA} architecture of the arrays at both ends of the system, we propose a novel multi-slot sensing strategy for \ac{UE} parameter estimation.
    \item We provide numerical results to demonstrate the effective gain resulting from increasing the physical size of the \ac{HIRS} array.
\end{itemize}



\subsubsection*{Notation} We adopt the following standard notation. $(\cdot)^*$ and $(\cdot)^\transp$ denote the complex conjugate and transpose operations, respectively.  $(\cdot)^\herm$ denotes the Hermitian (conjugate and transpose) operation. $\left|x\right|$ denotes the absolute value of $x$ if $x\in\RR$, while $|\Xc|$ denotes the cardinality of a set $\Xc$.  $\|\xv\|_2$ denotes the $\ell_2$-norm of a complex or real vector $\xv$. 
$\Id_m$ denotes the $m \times m$  identity matrix and $[n]=\{1, \dots, n\}$ the set of positive integers.  

\section{System Model}
\begin{figure}[h]
    \centering
  	\includegraphics[width=.4\textwidth]{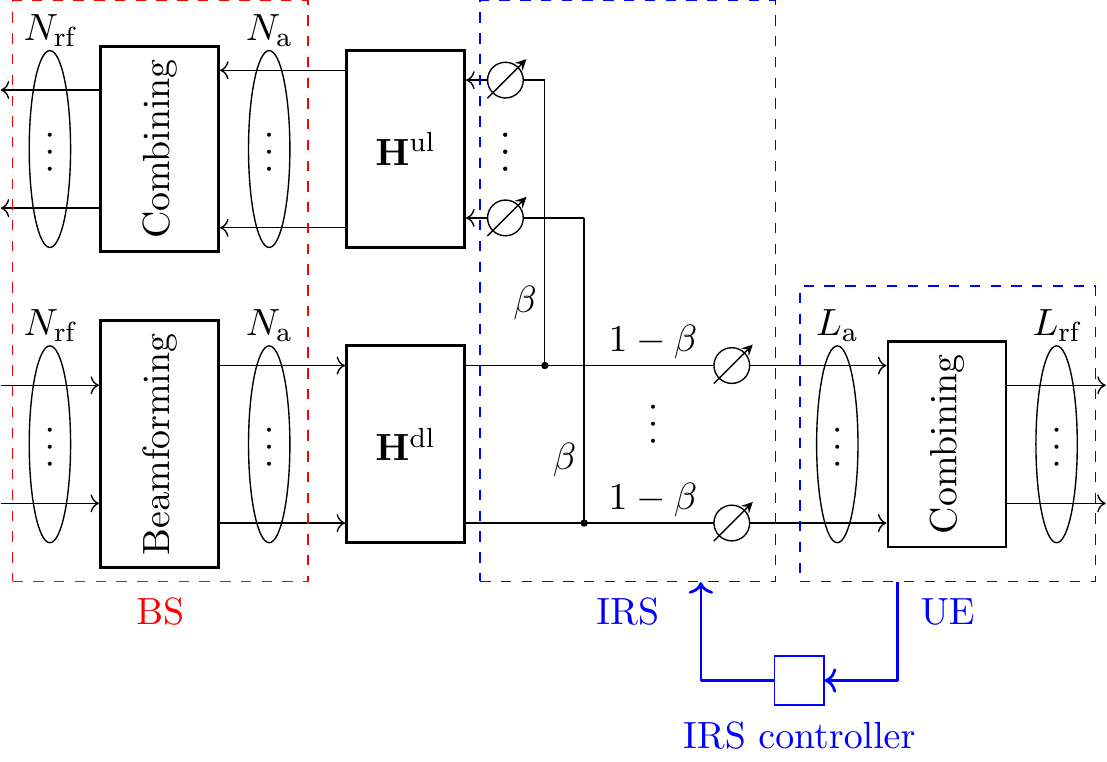}
  	\caption{Schematic of the system model. The \ac{HIRS} architecture adopts the model in \cite{alexandropoulos2021hybrid}.}
  	\label{fig:system_schematic}
\end{figure}

Consider a \ac{BS} and a \ac{UE} that is equipped with an \ac{HIRS} \cite{alexandropoulos2021hybrid}, as depicted in Fig. \ref{fig:system_schematic}. The \ac{BS} has $\Na$ antennas and $\Nrf$ \ac{RF} chains, while the \ac{HIRS} at the \ac{UE} side has $\La$ antennas, namely the $\La$ surface elements of the \ac{HIRS}, and $\Lrf$ \ac{RF} chains. The \ac{UE} is connected to the \ac{IRS} controller that performs \ac{BA}. An \ac{HIRS} can sense a portion of the incoming signal and reflect the remaining part in a controllable direction\cite{Alamzadeh2021}. 
For an incident signal $\xv \in \CC^{\La}$, the reflection and sensing signals are $\Phim^\herm \xv$ and $\Dm^\herm \xv$, respectively, where 
\begin{align}\label{eq:refl_matrix}
    \Phim &= \diag\left(\beta_1 e^{j \psi_1}, \dots, \beta_{l} e^{j \psi_{l}}, \dots, \beta_{\La} e^{j \psi_{\La}}\right)\\ \label{eq:sens_matrix}
    \Dm &= \diag\left(\bar{\beta}_1 e^{j \rho_1}, \dots, \bar{\beta}_l e^{j \rho_{l}}, \dots, \bar{\beta}_{\La} e^{j \rho_{\La}}\right)
\end{align}
are $\La\times\La$ complex reflection and sensing matrices, and where for $l \in [\La]$ the parameter $\beta_l \in [0,1]$ is the amplitude of the reflection coefficients, $\psi_l \in [-\pi, \pi]$ is the tunable phase shift of the reflected signal, $\rho_l \in [-\pi, \pi]$ is the tunable phase shift for the sensed signal and $\bar{\beta}_l = 1 - \beta_l$. 

We assume that the phase shifts can be compensated at the combining stage of the \ac{UE} and thus we set $\rho_l=0\ \forall\ l$ in \eqref{eq:sens_matrix}. For simplicity, we choose $\beta = \beta_1 = \beta_2 = \dots = \beta_{\La}$ so that
\begin{align}
    \Phim(\beta, \psi) &= \beta\;\diag\left(e^{j \psi_l}, \dots, e^{j \psi_{\La}}\right) \in \CC^{\La \times \La}\\ 
    \Dm(\beta) &= (1-\beta)\; \Id_{\La}  \in \CC^{\La \times \La}
\end{align}

\subsection{Channel Model}
Suppose the \ac{BS} and the \ac{UE} are equipped with \acp{ULA} with half-wavelength spacings (i.e. $\lambda_c/2$) between the antenna elements. The array response vectors at the \ac{BS} and \ac{UE} are denoted by
\begin{align}
    [\av(\theta)]_m &= e^{j \pi (m-1) \sin(\theta)} \quad m \in [\Na]\\
    [\bv(\phi)]_l &= e^{j \pi (l-1) \sin(\phi)} \quad l \in [\La],
\end{align}
where $\theta \in [-\frac{\pi}{2}, \frac{\pi}{2}]$ is the \ac{AoA} or \ac{AoD} at the \ac{BS}, and $\phi \in [-\frac{\pi}{2}, \frac{\pi}{2}]$ is the \ac{AoA} or \ac{AoD} at the \ac{UE}. 


For the \ac{DL} and \ac{UL} transmission, a linear time-varying \ac{LOS} channel is considered. In the delay-Doppler domain, the \ac{DL} and \ac{UL} channels are
\begin{align}\label{eq:dl_channel}
    \Hm^{\rm dl}(\tau,\nu) &= h^{\rm dl} \bv(\phi)\av^\transp(\theta) \delta(\tau - \tau_0/2)\delta(\nu - \nu_0/2) \in \CC^{\La\times\Na}\\\label{eq:ul_channel}
    \Hm^{\rm ul}(\tau,\nu) &= h^{\rm ul} \av(\theta)\bv^\transp(\phi) \delta(\tau - \tau_0/2)\delta(\nu - \nu_0/2) \in \CC^{\Na\times\La}
\end{align}
where $h^{\rm dl}$ and $h^{\rm ul}$ are the attenuation coefficients, $\tau_0$ is the two-way delay and $\nu_0$ is the two-way Doppler shift. 

The overall two-way channel $\Hm_i(\tau, \nu) \in \CC^{\Na \times \Na}$ in the $i$-th slot can be written as a two-dimensional convolution as
\begin{align}\nonumber
    \Hm_i(\tau, \nu) &= \Hm^{\rm ul} (\tau,\nu) \ast \Phim^\herm_i \Hm^{\rm dl}(\tau,\nu)\\ \nonumber
    &= h^{\rm dl} h^{\rm ul} \av(\theta) \bv^\transp(\phi)\Phim_i^\herm \bv(\phi) \av^\transp(\theta)\delta(\tau - \tau_0) \delta(\nu - \nu_0)\\
    &= h(\Phim_i) \av(\theta) \av^\transp(\theta) \delta(\tau - \tau_0) \delta(\nu - \nu_0),
    \label{eq:two_way_channel}
\end{align}
where $\Phim_i$ is the reflection matrix of the \ac{IRS} configured by the \ac{UE} in the $i$-th slot and
\begin{align}\label{eq:two-way-coeff}
    h(\Phim_i) \coloneqq h^{\rm dl} h^{\rm ul} \bv^\transp(\phi) \Phim^\herm_i \bv(\phi).
\end{align}
is the two-way channel coefficient. 

\subsection{Orthogonal  Frequency-Division  Multiplexing Signaling} 
We consider multi carrier modulation with \ac{OFDM}. To avoid \ac{ISI} between \ac{OFDM} symbols, each symbol is preceded by a \ac{CP} of duration $\Tcp$, resulting in an overall symbol duration of $T_o = T + \Tcp$. The \ac{OFDM} modulated signal in the $i$-th slot is thus
\begin{align}\label{eq:OFDM}
    s_{i}(t) = \sum_{n,m} x_i[n,m] \rect \left(\tfrac{t- n T_o}{T_o}\right) e^{j 2 \pi m \Delta f (t - \Tcp - n T_o)}
\end{align}
with average power constraint
\begin{align*}
    \EE[|x_i[n,m]|^2] = P_{\rm t}, \quad \forall (i,n,m)
\end{align*}
We assume that pilot symbols are transmitted from the \ac{BS} for the entire \ac{BA} duration. For simplicity, we consider a single stream \ac{DL} transmission such that we can express the beamformed transmitted signal as
\begin{align}\label{eq:OFDM_2}
    \sv_{i}(t) = \fv \sum_{n,m} x_i[n,m] \rect\left(\tfrac{t- n T_o}{T_o}\right) e^{j 2 \pi m \Delta f (t - \Tcp - n T_o)}
\end{align}
where $\fv \in \CC^{\Na}$ is a generic \ac{BF} vector of unit norm. We design $\fv$ so that it covers a section of the beam-space with a constant gain in the main beam, and very low gain elsewhere (see \cite{OTFS_MIMO_AV_22} for details).

\subsection{Received Signal Model}\label{rx_sig_mod}

The received signal at the \ac{UE} after channel \eqref{eq:dl_channel} is processed by the sensing matrix $\Dm_{i}$ and a combining matrix $\Um_{i}\in\CC^{\La\times\Lrf}$ resulting in the analog linear processing matrix $\Vm_{i} = \Dm_{i}\Um_{i}$. After removing the \ac{CP} and applying standard \ac{OFDM} processing, the sampled signal is (see e.g. \cite{sturm2011waveform})
\begin{align}
	\yv_i&[n, m] =\nonumber \\
	& g^\textrm{dl} \Vm_i^\herm \bv(\phi) x_i[n,m] e^{j2\pi (nT_o\frac{\nu_0}{2} -m \Delta f\frac{\tau_0}{2}}) + \wv_i[n,m],
	\label{eq:y_ue_after_dft}
\end{align}
where we have defined $g^\textrm{dl}\coloneqq h^{\rm dl}\av^\transp(\theta)\fv$.
Similarly, the received (back-scattered-) signal at the \ac{BS} at the $i$-th slot is
\begin{align}
	\rsf_i&[n, m] = \nonumber\\ 
	& g_i^\textrm{ul} \Um_{\rm BS,i}^\herm \bv(\theta) x_i[n, m] e^{j2\pi(nT_o\nu_0 -m\Delta f \tau_0)} + \nv_i[n, m],
	\label{eq:r_bs_after_dft}
\end{align}
where $\nv_i[n, m]=\frac{1}{M}\sum_{k=0}^{M-1} \nv_i[n,k] e^{-j2\pi \frac{mk}{M}}$ is the noise after the \ac{DFT} with $\nv_i[n,m] \sim \Nc_{\Cc}(\zerov, \sigma^2 \Id_{\Nrf})$, and $g_i^\textrm{ul} \coloneqq h(\Phim_i) \bv^\T(\phi) \fv$ the overall complex \ac{UL} channel coefficient. 
\subsection{Design of Receive Beamformers} 
As discussed in section \ref{rx_sig_mod}, the \ac{UE} and \ac{BS} apply a combining matrix to the signal received at their respective \acp{ULA}, due to the implemented hybrid BF architecture. To meet the page limit in this article, we provide only a brief overview of the design strategy for the sequence of combining matrices. The main concept here, is to design these matrices such that they probe different narrow angular sectors of
the beam space across different slots. To this end, using a method based on solving a magnitude least-squares problem for designing \ac{BF} vectors in \cite{OTFS_MIMO_AV_22, Dehkordi_OTFS_ICC}, we obtain a codebook of beamforming vectors $\mathcal{U}_{\text{UE}} = \{\uv_1,...,\uv_K\}$, where each of the $k\in [K]$ codewords is a \textit{flat-top} beam designed to cover a specific section on the desired field of view such that the codewords are not overlapping. In every slot $i$ of \ac{BA}, the \ac{UE} randomly samples $\Lrf$ BF vectors $\{\bar{\uv}_1,...,\bar{\uv}_{\Lrf}\}$ from~ $\mathcal{U}_{\text{UE}}$ and obtains its combining matrix, i.e. $\Um_{i} = \frac{1}{\sqrt{\Lrf}}\{\bar{\uv}_1,...,\bar{\uv}_{\Lrf}\}$ .  A similar procedure takes place at the \ac{BS} to obtain the \ac{BF} vectors $\Um_{\text{BS},i}$ indicated in \eqref{eq:r_bs_after_dft}.

\section{Beam Alignment}

\subsection{Multi-Slot Maximum Likelihood Estimation}\label{ssec:ML_est}
To solve the \ac{BA} problem, both the \ac{BS} and \ac{UE} must estimate their \acp{AoA}. We derive a \ac{ML} scheme and, to increase the accuracy, we suppose that in a certain slot of \ac{BA} all the observations up to the current slot are taken into account for the \ac{AoA} estimation so that the accuracy improves over time. 
Since the overall complex \ac{UL} channel coefficient $g_i^\textrm{ul}$ of the \ac{BS} might vary in each slot due to the chosen \ac{IRS} configuration, we additionally derive the \ac{MMLE} at the \ac{BS} for the case of slot-wise varying complex channel coefficients.
We thus first derive the multi-slot \ac{ML} estimate at \ac{UE} and we rewrite \eqref{eq:y_ue_after_dft} as
\begin{align}\label{eq:UE_rx_est}
    \ysf_i[n,m] = g^\textrm{dl} \Vm_i^\herm \bv(\phi) x_i[n,m] t_{n,m}(\tau_0,\nu_0) + \wsf_i[n,m],
\end{align}
where $t_{n,m}(\tau_0,\nu_0) \coloneqq e^{-j \pi \left(m \Delta f \tau_0 -  n T_o\nu_0\right)}$. We can reformulate the expression of \eqref{eq:UE_rx_est} by stacking the $N M$ observations into a column vector $\ysf_i \in \CC^{N M \Lrf}$ and, defining the expression $\Gm_i(\tau_0, \nu_0, \phi) \coloneqq \left(\Tm(\tau_0,\nu_0) \otimes \Vm_i^\herm \bv(\phi) \right)$, the column vector can be defined as
\begin{align}
    \ysf_i = g^\textrm{dl} \Gm_i(\tau_0, \nu_0, \phi) \xsf_i + \wsf_i.
\end{align}
The likelihood-function of $\ysf_i$ is
\begin{align} \label{eq:UE_likelihood}
    L(\ysf_i; (g^\textrm{dl},\tau_0,\nu_0,\phi)) = \frac{1}{\det(2\pi\sigma^2\Id_{NM\Lrf})^{1/2}} \cdot \nonumber \\
	\exp\left(-\frac{1}{2\sigma^2}\left( (\ysf_i - g^\textrm{dl} \Gm_i \xsf_i)^\herm (\ysf_i - g^\textrm{dl} \Gm_i \xsf_i)  \right) \right).
\end{align}
After collecting all the previous observations up to the $i$-th slot $\ysf^{(i)} =[\ysf_1, \dots, \ysf_i]$, the log-likelihood function is
\begin{align} \label{eq:UE_log_likelihood}
    \ell (\ysf^{(i)}; & (g^{\rm dl}, \tau_0, \nu_0, \phi)) = \log \left( L(\ysf^{(i)}; (g^\textrm{dl},\tau_0,\nu_0,\phi))\right) \nonumber \\
    &= \sum_{s=1}^{i} \log \left( L(\ysf_i; (g^\textrm{dl},\tau_0,\nu_0,\phi)) \right).
\end{align}
Using the \ac{ML} estimates for unknown parameters in \cite{scharf1991statistical} and 
\vspace{-0.6cm}
\begin{align}\nonumber
    \Vm_{(i)} &= \sum_{s=1}^{i} \lVert \xsf_s \rVert_2^2 \Vm_s \Vm_s^\herm\\
	\cv_{(i)}(\tau_0,\nu_0) &= \left[ \sum_{s=1}^{i} \xsf_s^\transp \Tm(\tau_0, \nu_0) \Ym_s^\herm \Vm_s^\herm\right] \nonumber,
\end{align}
we can write the \ac{ML} estimate as 
\begin{align}\label{eq:UE_estimate}
    \begin{split}
        (\hat{g}_i^\textrm{dl}, \hat{\tau}_i, \hat{\nu}_i, \hat{\phi}_i) = \argmax_{g^\textrm{dl}, \tau_0, \nu_0, \phi} \Re \left\{ 2 g^\textrm{dl} \cv^\herm_{(i)}(\tau_0,\nu_0) \bv(\phi) \right. \\
        \left. - \big| g^\textrm{dl} \big|^2 \bv^\herm(\phi) \Vm_{(i)} \bv(\phi)  \right\}
    \end{split}
\end{align}
Optimizing \eqref{eq:UE_estimate} with the respect of $\Re(g^\textrm{dl})$ and $\Im (g^\textrm{dl})$, we obtain
\vspace{-0.3cm}
\begin{align}
    	g_\textrm{opt}^\textrm{dl} = \frac{\bv^\herm(\phi) \cv_{(i)}(\tau_0,\nu_0)}{\bv^\herm(\phi) \Vm_{(i)} \bv(\phi)},
\end{align}
and the \ac{ML} estimates
\begin{align} \label{eq:UE_ml_estimate_opt}
    (\hat{\tau}_i, \hat{\nu}_i, \hat{\phi}_i) = \argmax_{\tau_0, \nu_0, \phi} \frac{\left\lvert \bv^\herm(\phi) \cv_{(i)}(\tau_0,\nu_0) \right\rvert^2}{\bv^\herm(\phi) \Vm_{(i)} \bv(\phi)},
\end{align}
which are approximately found by evaluating the objective function in a finite set of points.

At the \ac{BS} we use the same steps except that the channel coefficients in \eqref{eq:r_bs_after_dft} depend on the slot index. We can thus rewrite the received signal \eqref{eq:r_bs_after_dft} at the \ac{BS} as
\begin{align}
    \rsf_i = \left( \tilde{\Tm}(\tau_0, \nu_0) \otimes g_i^{\rm ul} \Um^\herm_{\rm BS, i}\av(\theta)\right)\xsf_i + \nv_i
\end{align}
Hence, the \ac{ML} estimate $(\{\hat{g}^\textrm{ul}\}_{s=1}^i, \hat{\tau}_i, \hat{\nu}_i, \hat{\theta}_i )$ is
\begin{align}\label{eq:BS_estimate}
\begin{split}
\hspace{-0.3cm}
    	\argmin_{\{g^\textrm{ul}\}_{s=1}^i, \tau_0, \nu_0, \theta} \Re \left\{ \sum_{s=1}^i \lvert g_s^{\rm ul}\lvert^2 \lVert \xsf_s \lVert_2^2 \av^\herm(\theta) \Um_{\rm BS,s} \Um^\herm_{\rm BS,s}\av(\theta)\right.\\
    	\left. - 2 g_s^{\rm ul} \xsf^\transp_s \tilde{\Tm}(\tau_0, \nu_0) \Rm_s^\herm \Um^\herm_{\rm BS, s} \av(\theta) \right\},
\end{split}
\end{align}
where $\Rm_s \in \CC^{\Nrf \times N M}$ is the matrix of the observation at \ac{BS} in the $s$-th slot.
As for the \ac{UE}, by defining 
\begin{align}\nonumber
    \tilde{\Um}_s &= \lVert \xsf_s \rVert_2^2 \Um_{\rm BS,s} \Um^\herm_{\rm BS,s} \\
    \tilde{\cv}_s(\tau_0,\nu_0) &= \left[ \xsf_s^\transp \tilde{\Tm}(\tau_0,\nu_0) \Rm_s^\herm \Um^\herm_{\rm BS,s} \right]^\herm,
\end{align}
and optimizing (\ref{eq:BS_estimate}) with the respect of $\Re(g_s^\textrm{ul})$ and $\Im (g_s^\textrm{ul})$, the optimal value of $g_s^\textrm{ul}$ is
\begin{equation}
	g_{s,\textrm{opt}}^\textrm{ul} = \frac{\av^\herm(\theta) \tilde{\cv}_s(\tau_0,\nu_0)}{\av^\herm(\theta) \tilde{\Um}_s \av(\theta)},
\end{equation}
which yields the \ac{ML} estimates at the \ac{BS} in the $i$-th slot as
\begin{equation} \label{ML3D}
    (\hat{\tau}_i, \hat{\nu}_i, \hat{\theta}_i) = \argmax_{\tau_0, \nu_0, \theta} \sum_{s=1}^i \frac{\lvert \av^\herm(\theta) \tilde{\cv}_s(\tau_0,\nu_0) \lvert^2}{\av^\herm(\theta) \tilde{\Um}_s \av(\theta)},
\end{equation}
which is approximately solved by evaluating the objective function on a finite set of points.

\subsection{Cramer Rao Lower Bound}
We derive the \ac{CRLB} as a benchmark. Let $g = |g^{\rm dl}|$ and $\psi_g = \angle(g^{\rm dl})$ be the amplitude and phase of $g^{\rm dl}$, respectively, and define the vector $\xiv = [g, \psi_g, \phi, \tau'_0, \nu_0]$ with the unknown real parameters. We form the $5\times 5$ Fisher information matrix whose $(k,l)$-th element is
\begin{align}
	[&\Id(\xiv, \Xm)]_{k,l} = \nonumber\\
	&\frac{2}{\sigma^2} \sum_{s=1}^{i} \sum_{n,m} \Re \left\{ \frac{\partial \sv_s^\herm[n,m;\xiv] }{\partial \xi_k} \frac{\partial \sv_s[n,m;\xiv] }{\partial \xi_l} \right\},
	\label{eq:fim_formula_mv_gaussian}
\end{align}
%
%
where $\Xm = \{ \Xm_1, \dots, \Xm_i \}$ is the set of all pilot symbols sent up to the $i$-th slot, and $\Xm_s = \{x_s[n,m]\}\ \forall\ n,m$ the set of all pilot symbols sent in the $s$-th slot. The expression \eqref{eq:fim_formula_mv_gaussian} can be manipulated to take the following structure:   
\begin{equation}
	\Id(\xiv, \Xm) 
	= \frac{1}{\sigma^2} \begin{bmatrix}
		I_{g g} & 0 & I_{g \phi} & 0 & 0 \\
		0 & I_{\psi_g \psi_g} & I_{\psi_g \phi} & I_{\psi_g \tau'_0} & I_{\psi_g \nu_0} \\
		I_{g \phi} & I_{\psi_g \phi} & I_{\phi \phi} & I_{\phi \tau'_0} & I_{\phi \nu_0} \\
		0 & I_{\psi_g \tau'_0} & I_{\phi \tau'_0} & I_{\tau'_0 \tau'_0} & I_{\tau'_0 \nu_0} \\
		0 & I_{\psi_g \nu_0} & I_{\phi \nu_0} & I_{\tau'_0 \nu_0} & I_{\nu_0 \nu_0} 
	\end{bmatrix},
	\label{eq:fim_simplified} 
\end{equation} 
Let $\hat{\phi}$ be an unbiased estimator of $\phi$. Since we consider only \ac{AoA} estimation, it can be further simplified to yield the approximated \ac{CRLB} in the $i$-th slot as \eqref{eq:crlb_phi_approx_simplified}, where we defined $\tilde{\bv}(\phi)$ as $\tilde{\bv}(\phi) = \diag(0,\dots,\La-1)\, \bv(\phi)$. 

\newcounter{TempEqCnt}
\setcounter{TempEqCnt}{\value{equation}}
\setcounter{equation}{28}
\begin{figure*}[ht]
\begin{align}
	\mathrm{Var}\{ \hat{\phi} \}_i &\geq \frac{ C^{(i)}_{\phi} \sigma^2 } 
	{ 2M N \Pav g^2 \pi^2 \cos^2(\phi) \left(  C^{(i)}_{\phi} \tilde{\tilde{C}}^{(i)}_{\phi} - \left( \Re \left\{ \tilde{C}^{(i)}_{\phi} \right\} \right)^2 \left[ 3\cdot(1-\cos(\phi))^2 + 1 \right] - \left( \Im \left\{ \tilde{C}^{(i)}_{\phi} \right\} \right)^2 \right)}\,, ~\label{eq:crlb_phi_approx_simplified}\\
	&\text{where}~~  C^{(i)}_{\phi} \coloneqq \sum_{s=1}^{i}\| \Vm_s^H\bv(\phi)\|_2^2~,~ \tilde{C}_{\phi}^{(i)} \coloneqq \sum_{s=1}^{i} \tilde{\bv}^H(\phi) \Vm_s \Vm_s^H \bv(\phi)~,~ \tilde{\tilde{C}}_{\phi}^{(i)} \coloneqq \sum_{s=1}^{i} \left\lVert \Vm_s^H \tilde{\bv}(\phi) \right\rVert_2^2 \nonumber.
\end{align}
\hrulefill
\end{figure*}

\subsection{\ac{IRS} parameter tuning}
We present here a method to set the \ac{IRS} parameters, namely $\beta$ and $\{\psi_{i}\}_{i=1}^{\La}$, in order to help the \ac{BS} estimate its \ac{AoD}. We define the moving standard deviation of the \ac{UE} local estimate at time slot $i$ as 
\begin{align}\label{eq:mov_st}
    \sigma(\phi_{i-N_{\rm w}+1}^{i}) \coloneqq \sqrt{\frac{1}{N_{\rm w}-1}\sum_{j=0}^{N_{\rm w}-1}\left(\hat\phi_{i-j} - \overline{\phi}_{i-N_{\rm w}+1}^{i}\right)^{2}},
\end{align}
where $\overline\phi_{i-N_{\rm w}+1}^{i} \coloneqq \frac{1}{N_{\rm w}}\sum_{j=0}^{N_{\rm w}-1}\hat{\phi}_{i-j}$ is the moving average of the estimate. Our method sets $\beta=0$ until the moving standard deviation drops below a predefined threshold, and $\beta=1$ thereafter. In particular, we select the threshold as the \SI{3}{\decibel} beamwidth of an $\La$-antenna \ac{ULA}, given by \cite[Ch. 6]{balanis2012antenna}
\begin{align}
    {\Theta}_{\SI{3}{\decibel}} = 2\left[\frac{\pi}{2} - \arccos{\frac{2\cdot1.391}{\pi \La}}\right].
\end{align}

Regarding the \ac{IRS} phase shifts, it is trivial to observe that the magnitude of the two-way coefficient in \eqref{eq:two-way-coeff} is maximized when we set $\psi_{i} = 2\pi(i-1)\sin(\phi)$. 

The resulting \ac{IRS} configuration strategy is
\begin{align}
    \begin{cases}
        \begin{aligned}
            \Phim_{i}(\beta, \psi) &= \zerov_{\La\times\La}\\
            \Dm_{i}(\beta) &= \Id_{\La}
        \end{aligned}
        &{\rm if}\quad{ \sigma(\phi_{i-N_{\rm w} + 1}^{i}}) > \Theta_{\SI{3}{\decibel}},\\[12pt]
        \begin{aligned}
            \Phim_{i}(\beta, \psi) &= \diag(\bv(2\hat\phi_{i}))\\
            \Dm_{i}(\beta) &= \zerov_{\La\times\La}
        \end{aligned}
        &{\rm if}\quad{ \sigma(\phi_{i-N_{\rm w} + 1}^{i}}) < \Theta_{\SI{3}{\decibel}},\\
    \end{cases}
\end{align}
where $\hat\phi_{i}$ is is the \ac{ML} estimate obtained by the \ac{UE} as described in section \ref{ssec:ML_est}.

\subsection{Radar Cross Section}
To model the two-way channel between \ac{BS} and \ac{UE},  it is fundamental to consider the \ac{RCS} of the \ac{IRS}. In each slot $i$ of \ac{BA}, the \ac{RCS} of the \ac{IRS} can be computed by
\begin{align}\label{eq:rcs_1}
    \sigma_{\rm RCS, i} \coloneqq \sigma_{\rm RCS, BBF} \cdot \cos(\phi) \cdot G_{\rm IRS}(\Phim_i)
\end{align}
where  $\sigma_{\rm RCS, BBF}$ denotes the \ac{RCS} of the \ac{IRS} before \ac{BF}. Given that the \ac{IRS} is configured for reflection towards a certain direction, its \ac{RCS} increases towards this direction by the achievable \ac{IRS} gain which is defined in (\ref{eq:two-way-coeff}) as 
\begin{align}\label{eq:irs_gain}
    G_{\rm IRS}(\Phim_i) \coloneqq  |\bv^\transp(\phi) \Phim^\herm_i \bv(\phi)|.
\end{align}
We propose a model for the \ac{RCS} of the \ac{IRS} based on \cite{https://doi.org/10.48550/arxiv.2211.08626}. This model is obtained by considering a realistic \ac{IRS} array composed of conventional metallic patches. This model takes into account the physical array dimensions and the operating wavelength. The numerical value is given by
\begin{align}\label{eq:rcs_2}
    \sigma_{\rm RCS, BBF} = \frac{4 \pi (\frac{\lambda_c}{2} \La)^2(\frac{\lambda_c}{2})^2}{\lambda_c^2}.
\end{align}
For performance comparison purposes, we consider also a hypothetical value of \ac{RCS} before \ac{BF}. For this purpose, we assume the \ac{IRS} fits within a conventional mobile phone.
Measurements of the back of a human hand \cite{7461605} or other similar-sized objects \cite{8714559} show that one can obtain an average \ac{RCS} between -20 dBsm and -15 dBsm. However, since such objects present curved shapes and less radar reflectivity than \ac{IRS}, it is reasonable to assume that the monostatic \ac{RCS} of the \ac{IRS} should be higher than these values, yielding $\sigma_{\rm RCS, BBF} > - 15$ dBsm. Recent work on drones' \ac{RCS} \cite{9032332} found that a metallic object with an area of 128 mm $\times$ 53 mm (similar size to a mobile phone) results in a \ac{RCS} value of $\sigma_{\rm RCS, MP}(\lambda_c) = 13$ dBsm at a carrier frequency $f_c = 60$ GHz. We thus assume that $\sigma_{\rm RCS, ABF}$ after perfect \ac{BF} is upper bounded as $\sigma_{\rm RCS, ABF} \leq \sigma_{\rm RCS, MP}(\lambda_c) = 13$ dBsm.  To this end, we select the hypothetical value of \SI{-5}{dBsm}  for this comparison. This value is justified since the \ac{IRS} gain is upper bounded by
\begin{align}
    G_{\rm IRS}(\Phim_i) = \lvert \bv^\transp(\phi) \Phim_i^\herm \bv(\phi)\lvert \leq \La = 64 \equiv \SI{18}{\decibel},
\end{align}
which means $\sigma_{\rm RCS, i}$ can reach its upper bound of \SI{13}{dBsm} in case of $\phi = 0$ and perfect reflection.

\section{Numerical Results}
We now provide numerical results to verify the effectiveness of the methods proposed in the previous section. In the remainder, we consider the parameters shown in Table \ref{tab_param}. The channel parameters in \eqref{eq:dl_channel} and \eqref{eq:ul_channel} are assumed to remain constant over $N_{\rm slot}$ slots, defined as the maximum number of slots expected to be necessary for \ac{BA}. This is justified for moderate values of $N_{\rm slot}$ since the frame duration is approximately \SI{50}{\micro\second}. Some of the results are given as a function of the \ac{SNR} that would be obtained at the \ac{UE} in case no beamforming would be used at the transmitter nor at the receiver. We refer to this magnitude as the \ac{SNR} before beamforming (${\rm SNR}_{\rm UE, BBF}$), which is given by
\begin{align}
    {\rm SNR}_{\rm UE, BBF} \coloneqq \frac{\lambda_{\rm c}^2}{(4\pi d)^2}\frac{P_{\rm t}}{\sigma^2}.
\end{align}

\begin{figure}[h]
    \centering
    \includegraphics[width=.4\textwidth]{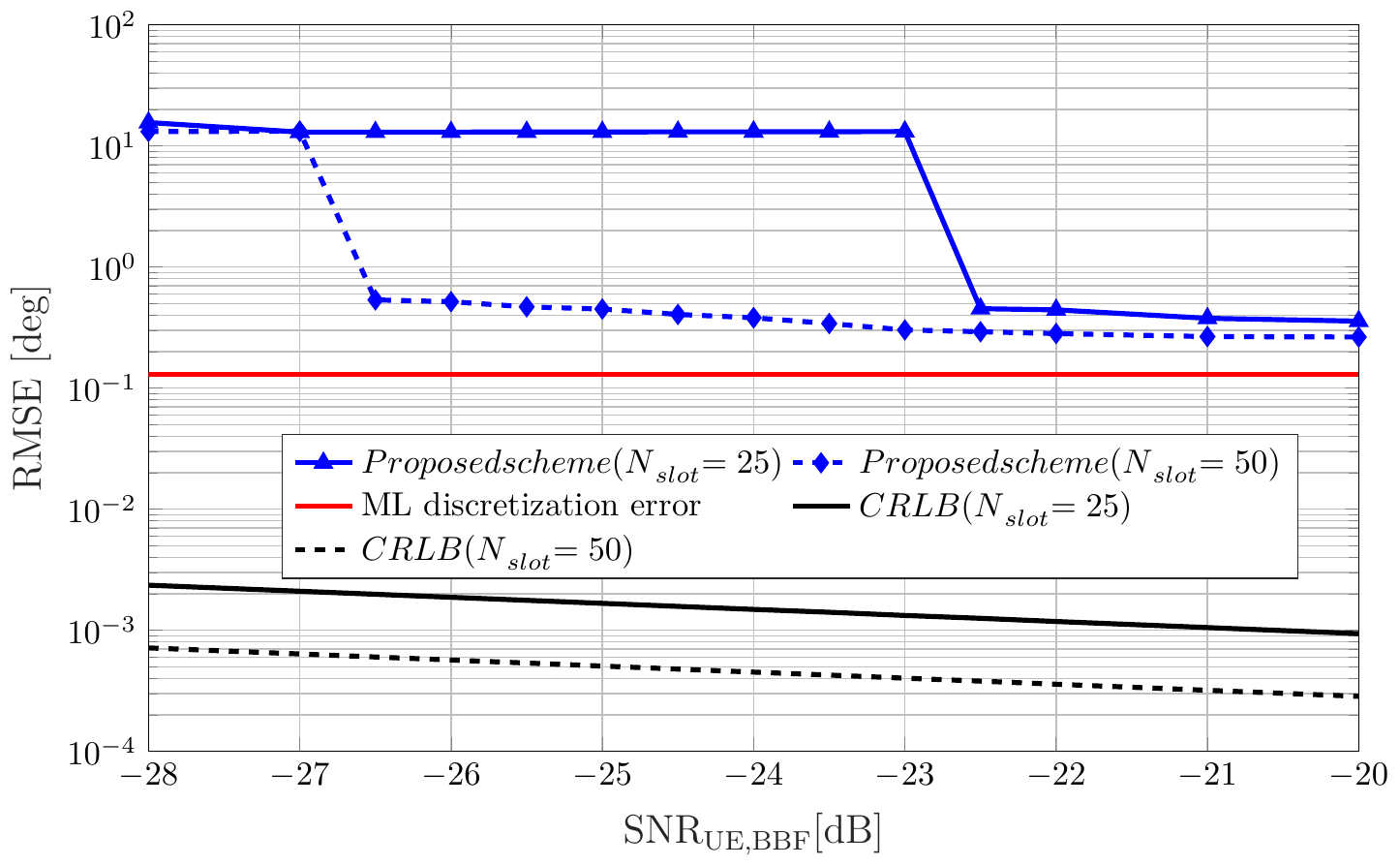}
  	\caption{RMSE value of the \ac{AoA} estimated at the user side. }
  	\label{fig:UE_aoa}
\end{figure}

First, the \ac{AoA} estimation accuracy at the \ac{UE} side is investigated. Figure~\ref{fig:UE_aoa} shows the estimated \ac{AoA} \ac{RMSE} as a function of SNR$_{\rm UE, BBF}$. For evaluation of the \ac{RMSE}, we run a large number of simulations over certain range of distances, where at each run, the \ac{AoA} and \ac{AoD} are chosen uniformly at random from the set $[-87^{\circ}, 87^{\circ}]$. Note that the discretization error of the ML estimation, i.e. the lowest achievable \ac{RMSE} due to the discretized grid for the ML estimation in (\ref{eq:UE_ml_estimate_opt}), is shown to evaluate the general quality of the \ac{MMLE} results. It can be observed that the proposed estimation scheme improves significantly with larger number of slots for \ac{BA}. We would like to further remark that, although the above simulation presents the \ac{ML} estimate of the \ac{AoA}, the \ac{ML} estimation metrics in (\ref{eq:UE_ml_estimate_opt}), and (\ref{ML3D}) at the \ac{UE} and \ac{BS} side respectively, can be used to obtain an estimate of the delay, Doppler and angle parameters simultaneously, where these parameters are defined over a 3-dimensional grid of parameters.  

The following figures indicate performance in terms of the achievable spectral efficiency at the \ac{UE} after obtaining angular estimates and using them to tune the beamformers.  This is numerically computed by averaging
\begin{align}
    \log_{2}\left(1+{\rm SNR}_{\rm UE, BBF}|\av^\transp(\theta)\av^*(\hat\theta)\bv^\herm (\hat\phi)\bv(\phi)|^2\right)
\end{align}
over multiple simulations over a range of distances, where $\hat\phi$ and $\hat\theta$ are \ac{ML} estimates obtained as derived in Section \ref{ssec:ML_est}.

\begin{figure}[h]
    \centering
  	\includegraphics[width=.4\textwidth]{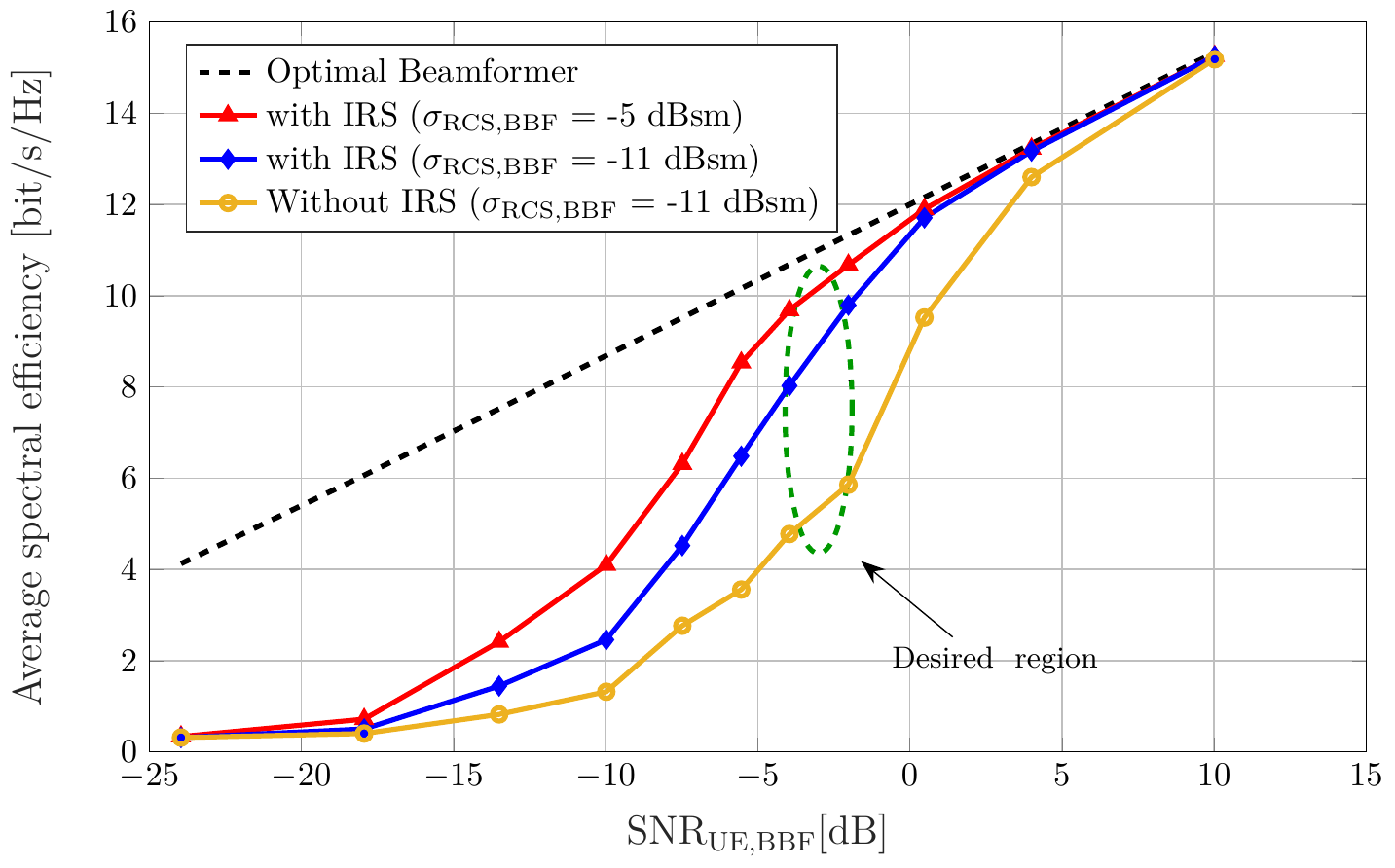}
  	\caption{Averaged spectral efficiency as function of \ac{UE} for different RCS values. }
  	\label{fig:spect_eff__var_SNR}
\end{figure}

It is easy to verify that, by applying the values in Table \ref{tab_param} to \eqref{eq:rcs_2}, the \ac{RCS} evaluates to approximately \SI{-11}{\deci\belsquaremeter}. The achievable spectral efficiency after beamforming when the number of slots for \ac{BA} is fixed to 32 is shown in Figure~\ref{fig:spect_eff__var_SNR}. There, we consider the analytic \ac{RCS}, a hypothetical one, and a case where the \ac{IRS} is replaced by a metallic plate of the same size. It can be observed that the communication performance after \ac{BA} is close to optimal for both \ac{RCS} values when the \ac{SNR} is as low as \SI{-5}{\decibel} to \SI{0}{\decibel}.

\begin{figure}[h]
    \centering
    \includegraphics[width=.4\textwidth]{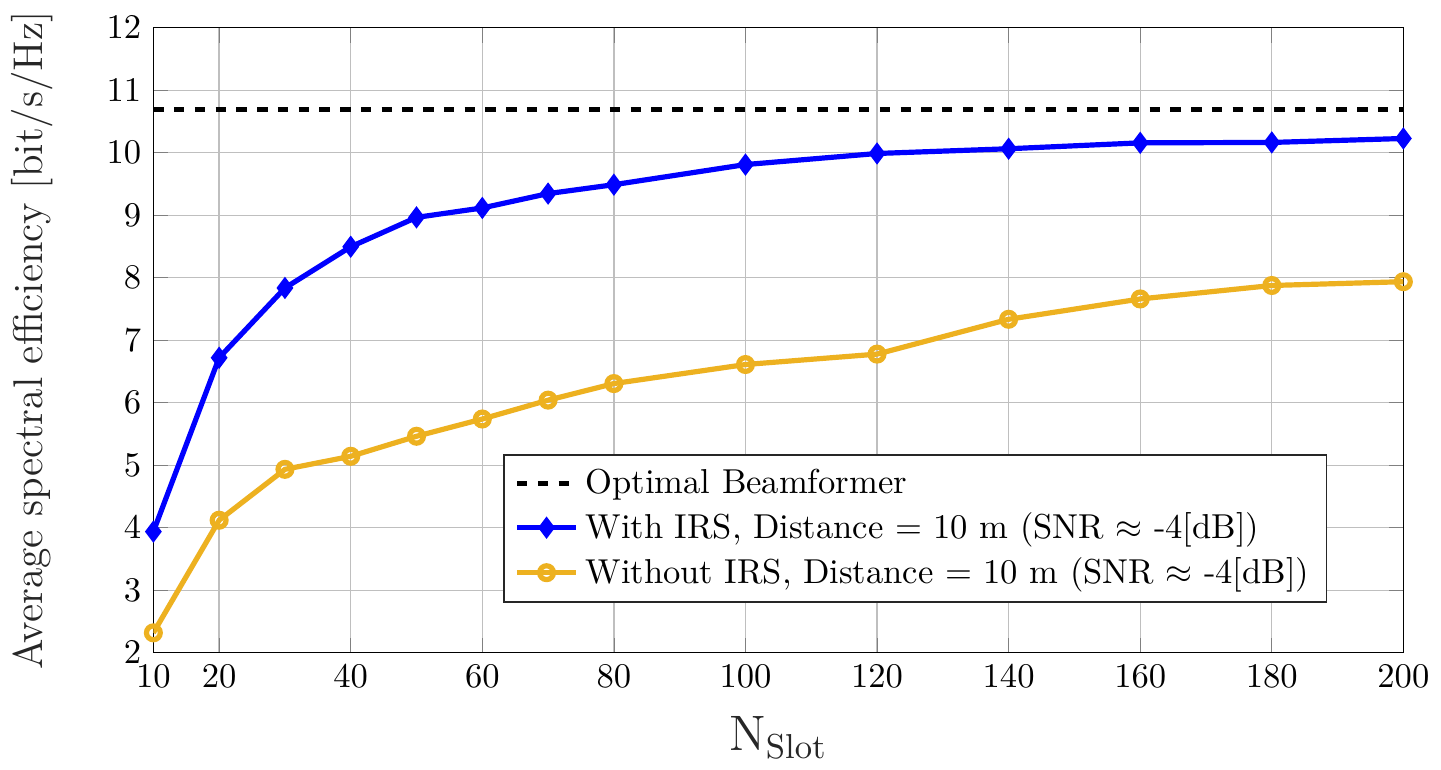}
    \caption{Achievable spectral efficiency as a function of number of estimation blocks for a fixed user at a distance of \SI{10}{m}. }
     \label{fig: se_slot}
\end{figure}

Inspired by the previous result, we now fix the ${\rm SNR}_{\rm UE,BBF}$ to \SI{-4}{\decibel} (corresponding to a distance of \SI{10}{\meter} for our system configuration, reasonable for indoor scenarios) and study performance in terms of achievable spectral efficiency as a function of the number of slots allocated for \ac{BA} in Figure~\ref{fig: se_slot}. The result shows that our \ac{IRS} based \ac{BA} method consistently improves spectral efficiency by at least \SI{2}{\bit/\second/\hertz}. Note that in all of the above simulations we have used a relatively small transmit power of $1$mW. By using larger values, as would be the case with most \acp{BS}, the effective operational range of the scheme can be extended to meet requirements for larger cell sizes.

\begin{table}\renewcommand{\arraystretch}{1.2} 
	\centering
	\begin{tabular}{>{\centering} m{0.24\textwidth} >{\centering\arraybackslash}m{0.2\textwidth}}
		\hline
		\textbf{Parameter} & \textbf{Value} \\
		\hline
		Operating frequency & $f_c = \SI{60}{\giga\hertz}$ $\Leftrightarrow$ $\lambda_{\textrm{c}} = \SI{5}{\milli\meter}$\\ 
		Bandwidth & $B \approx \SI{1}{\giga\hertz}$ \\ 
		Subcarriers & $M=2048$ \\ 
		Subcarrier-spacing & $\Delta f = \SI{480}{\kilo\hertz}$ \\ 
		\ac{OFDM} symbols per slot & $N=14$ \\ 
		\ac{CP} duration & $\Tcp = 0.07/{\Delta f}$ \\ 
		\ac{BS} antennas & $\Na = 64$ \\ 
		\ac{IRS} antennas/elements & $\La = 64$ \\ 
		\ac{RF} chains \ac{BS}/\ac{UE} & $\Nrf = \Lrf = 4$ \\ 
		Transmit power & $P_t = \SI{0}{dBm} = \SI{1}{\milli\watt}$\\ 
		Noise power & $\sigma^2 = \SI{-84}{dBm} \approx \SI{4e-12}{\watt}$ \\ 
		Pilot signals & similar to CSI-RS from \cite{3GPP-38_211} \\
        RCS model of \ac{IRS} & See \eqref{eq:rcs_1} and \eqref{eq:rcs_2} \\ 
		\ac{ML} grid size (angle, delay, Doppler) & $400 \times 20 \times 20$ \\
		No. of estimates for \ac{IRS} activation & $N_{\textrm{w}} = 5$ \\
		\hline \vspace{0.8mm}
	\end{tabular} 
	\caption{Overview of used system parameters.}
	\label{tab_param}
\end{table}\noindent


\section{Conclusions}
Motivated by the requirement for \ac{BA} in mmWave communications with highly directional beamforming, we have proposed the use of an on-device-mounted \ac{HIRS} to aid the \ac{BA} procedure. In the proposed scheme, a multi-slot parameter estimation framework is developed to deal with the restriction imposed by the \ac{HDA} architecture. Our numerical results demonstrate that with sufficiently large number of slots, the user device can reliably estimate the \ac{AoA} of the incoming communication signal and maintain a significantly higher spectral efficiency.

\section{Acknowledgment}
The authors would like to thank Gerhard Kramer for his careful reading of the manuscript and his very useful feedback.

S. K. Dehkordi and F. Pedraza  would like to acknowledge the financial support by the Federal Ministry of Education and Research of Germany in the program of “Souverän. Digital. Vernetzt.” Joint project 6G-RIC, project identification number: 16KISK030.

The research of Lorenzo Zaniboni is funded by Deutsche Forschungsgemeinschaft (DFG) through the grant KR 3517/12-1.

\bibliographystyle{IEEEtran}
\bibliography{radarcom,IEEEabrv}

\end{document}